\documentclass[reprint,twocolumn,showpacs,preprintnumbers,superscriptaddress,amsmath,amssymb,floatfix, nofootinbib,pra]{revtex4-1}

\usepackage{graphicx}
\usepackage{bm}
\usepackage{bbm}
\usepackage{bbold,color}
\usepackage{CJK}
\usepackage{feynmp}
\usepackage{comment}

\newcommand{\mri}{\mathrm{i}}

\newcommand{\Exp}[1]{\mathrm{e}^{\mbox{\footnotesize$#1$}}}

\begin{document}

\begin{CJK*}{UTF8}{gbsn}
\title{Dynamics of localized waves in 1D random potentials: statistical theory of the coherent forward scattering peak}

\author{Kean Loon \surname{Lee} (李健伦) }
\affiliation{Centre for Quantum Technologies, National University of Singapore, 3 Science Drive 2, Singapore 117543, Singapore}

\author{Beno\^it \surname{Gr\'emaud}}
\affiliation{Laboratoire Kastler-Brossel, UPMC-Paris 6, ENS, CNRS; 4 Place Jussieu, F-75005 Paris, France}
\affiliation{Centre for Quantum Technologies, %
National University of Singapore, 3 Science Drive 2, Singapore 117543, %
Singapore}
\affiliation{Department of Physics, National University of Singapore, %
2 Science Drive 3, Singapore 117542, Singapore}

\author{Christian \surname{Miniatura}}
\affiliation{Merlion MajuLab, UMI 3654, CNRS, UNS, NUS, NTU, Singapore}
\affiliation{INLN, Universit\'e de Nice-Sophia Antipolis, CNRS; 1361 route des Lucioles, 06560 Valbonne, France}
\affiliation{Centre for Quantum Technologies, %
National University of Singapore, 3 Science Drive 2, Singapore 117543, %
Singapore}
\affiliation{Department of Physics, National University of Singapore, %
2 Science Drive 3, Singapore 117542, Singapore}
\affiliation{Institute of Advanced Studies, Nanyang Technological University, %
60 Nanyang View, Singapore 639673, %
Singapore}

\pacs{05.60.Gg, 03.75.-b, 42.25.Dd, 72.15.Rn}

\begin{abstract}
{
As recently discovered [PRL {\bf 109} 190601(2012)], Anderson localization in a bulk disordered system triggers the emergence of a coherent forward scattering (CFS) peak in momentum space, which twins the well-known coherent backscattering (CBS) peak observed in weak localization experiments. Going beyond the perturbative regime, we address here the long-time dynamics of the CFS peak in a 1D random system and we relate this novel interference effect to the statistical properties of the eigenfunctions and eigenspectrum of the corresponding random Hamiltonian. Our numerical results show that the dynamics of the CFS peak is governed by the logarithmic level repulsion between localized states, with a time scale that is, with good accuracy, twice the Heisenberg time. This is in perfect agreement with recent findings based on the nonlinear $\sigma$-model. In the stationary regime, the width of the CFS peak in momentum space is inversely proportional to the localization length, reflecting the exponential decay of the eigenfunctions in real space, while its height is exactly twice the background, reflecting the Poisson statistical properties of the eigenfunctions. Our results should be easily extended to higher dimensional systems and other symmetry classes.
}
\end{abstract}

\begin{widetext}
\maketitle
\end{widetext}
\end{CJK*}

\section{Introduction}

Over the past decades, elucidating the interplay between multiple scattering and interference
has played a major role in our understanding of wave transport in disordered media, see \cite{akkermans2007, bergmann1984} and references therein. We now know that coherent corrections bring notable deviations to the usual classical diffusion transport theory as exemplified by weak localization corrections to the Boltzmann diffusion constant, universal conductance fluctuations in mesoscopic electronic systems, long-range intensity correlations in speckle patterns, or the celebrated coherent backscattering (CBS) effect~\cite{akkermans2007, albada1985, labeyrie1999, bidel2002}. Remarkably, interference inhibits transport and can ultimately bring it, under suitable conditions, to a complete stop, a phenomenon commonly known as Anderson (or strong) localization (AL)~\cite{anderson1958,kramer1993}. In fact, under the hypothesis of a one-parameter scaling, AL is the rule for one-dimensional (1D) and two-dimensional (2D) bulk systems while a disorder-induced metal-insulator transition takes place in three-dimension (3D)~\cite{abrahams1979}. AL has been actively studied with light waves~\cite{wiersma1997,sperling2013}, polaritons~\cite{cheng1990}, acoustic waves~\cite{hu2008,faez2009}, water waves~\cite{belzons1988}, ultracold atoms~\cite{lemarie2008,billy2008,roati2008,jendrzejewski2012,kondov2011}, and quantum Hall systems~\cite{ilani2004}. Its unambiguous experimental observation remains difficult, often controversial, as spurious effects like absorption, dephasing or nonlinear effects should be completely suppressed. At the same time, it should be clearly distinguished from other types of localization, such as the Mott-insulator transition~\cite{mott1949} or classical trapping in disconnected classically-allowed regions~\cite{pezze2011}.

Recently we have proposed to monitor AL for matter waves in momentum space~\cite{cherroret2012,karpiuk2012}. Indeed, observing the CBS effect \cite{cherroret2012} ensures that interference is at work and that phase coherence is preserved, while observing the CFS effect \cite{karpiuk2012} makes sure that the bulk system has entered the AL regime. To date, while detection and characterization of the CBS peak in momentum space have been quickly reported for matter waves~\cite{jendrzejewski2012_cbs,labeyriel2012}, the CFS peak still calls for an experimental observation. We propose to search for the CFS peak in a 1D speckle system as realized in~\cite{billy2008} for example.
Indeed, in 1D random systems, the localization length $\xi$ at a given energy $E$ scales linearly with the transport mean free path $\ell_B$ and AL is thus more easily accessible than in 2D systems (where $\xi$ scales exponentially with $\ell_B$) or in 3D systems (where a mobility edge exists and is difficult to reach). In particular moderate disorder strengths are sufficient to get localization lengths well below the system size.

In~\cite{karpiuk2012}, building on a perturbative diagrammatic theory, we have argued that the constructive interference of counter-propagating multiple scattering amplitudes traveling along loop-like paths in real space an even (resp. odd) number of times contribute to the CFS (resp. CBS) peak. We further suggested that the CFS peak grows in time with a timescale related to the Heisenberg time associated with the localization volume~\cite{karpiuk2012}. Unfortunately, but not surprisingly, the perturbative approach is not suited to tackle the long-time limit of the localization dynamics since it would require the resummation of the full diagrammatic series. To address the ultimate fate of the CBS and CFS peaks and make precise quantitative predictions on their shape, width and time dependence, one needs to resort to more powerful techniques. In this paper we use theoretical tools borrowed from random matrix theory~\cite{mehta2004}, as applied to the study of disordered quantum dots~\cite{altshuler1986,sivan1987,altshuler1989,prigodin1994,prigodin1995a,prigodin1995b,altland1995}, to analyze the statistical properties of the localized eigenstates and eigenvalues of our random Hamiltonian and to infer the properties of the CFS and CBS peaks. It is important to note that, if the statistical properties of localized eigenstates in real space have already been investigated in the literature, the interest and focus in this work lies in the statistical properties of the localized eigenstates in {\it momentum} space. In particular, we elucidate the quantitative connection between the CFS peak and the spatial as well as spectral correlations of the localized eigenstates. Importantly, our analysis can be generalized to higher-dimensional disordered systems, and to other symmetries classes than the one considered here, to further test the relationship between CFS and AL. More specifically, we show and explain that: \\

$\bullet$ the CFS peak dynamics is governed by the behavior of the auto-correlation function of the density of states per unit length (DOS) of the bulk system, which encapsulates the level repulsion induced by localized eigenstates located far apart in real space. Its characteristic time scale is twice the Heisenberg time. Noticeably this is also the underlying physics behind the low-frequency ac conductivity of a large disordered system~\cite{mott_davis1979,sivan1987};\\ 

$\bullet$ the height of the CFS peak in the stationary limit is exactly twice the diffusive background because of the Poisson statistics of the localized eigenstates;\\

$\bullet$ the shape of the CFS peak in the stationary limit is well approximated by the sum of a diffusive background and a Lorentzian function with a width given by the inverse of the localization length $\xi^{-1}$ because of the correlations of the localized eigenstates in momentum space.\\

These predictions, based on the statistical analysis of the spectral properties of our random Hamiltonian, apply irrespective of the disorder strength. We further corroborate them by numerically-solving the Schr\"odinger equation and extracting the various disorder-averaged quantities needed. Recently Micklitz {\it et al.}~\cite{micklitz2014} have investigated the CFS peak at a fixed energy for a quasi-1D system in the presence of a weak magnetic field. Using the supersymmetric nonlinear $\sigma$-model~\cite{efetov1983}, the Authors derived the time dependence of the CFS peak height and also concluded that the CFS peak is exactly twice the background in the long-time limit. While their results apply to the Gaussian Unitary Ensemble (GUE) and our system is described by the Gaussian Orthogonal Ensemble (GOE), we find surprisingly good agreement between our numerically-extracted time dependence of the CFS peak height and their analytical results.

The rest of the paper is organized as follows. In a first Section, we briefly describe our model random Hamiltonian, we compute numerically the disorder-averaged momentum distribution at different times and we relate it to the eigensystem of the Hamiltonian. In the following Section, we investigate the disorder-average momentum distribution in the long-time limit through the looking-glass of the statistical properties of the spectrum and the eigenstates. We compare our predictions to our numerical data. In a final Section, we turn to the time-dependence of the CFS peak contrast and its relation to the auto-correlator of the DOS. We conclude by mentioning possible future work.

\begin{figure*}
  \includegraphics[width=0.9\textwidth]{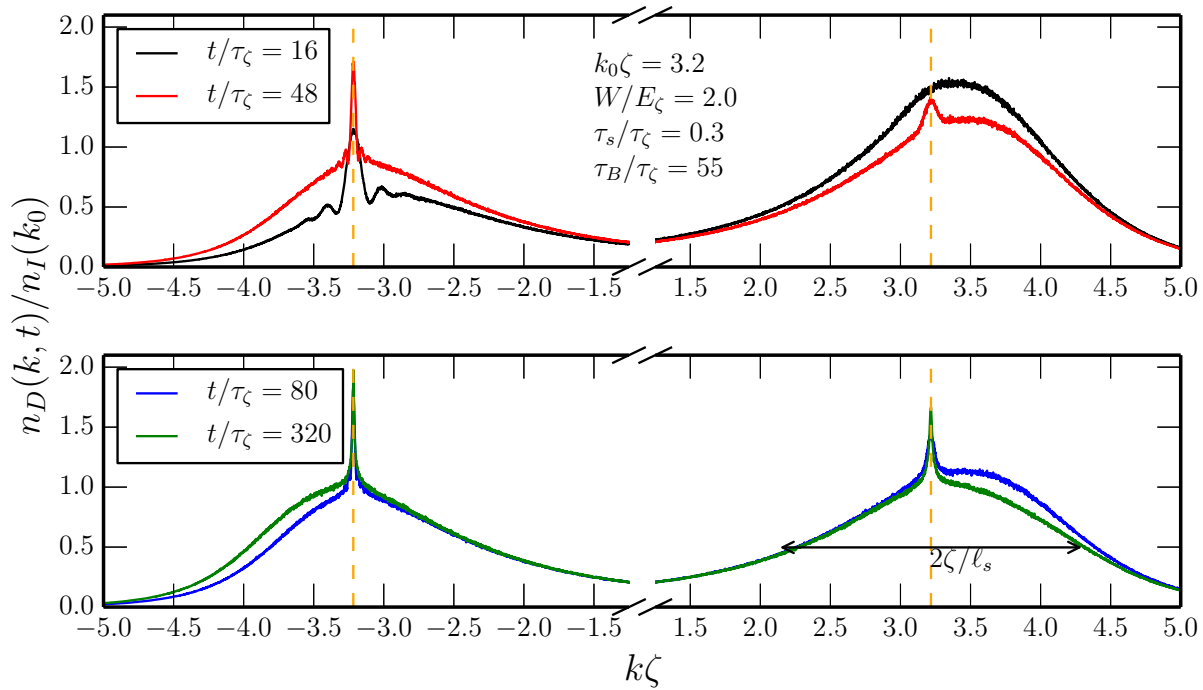}\caption{\label{fig:nk_dynamics_full} (Color online) Diffuse momentum distribution $n_D(k, t)$, normalized by its stationary incoherent background value $n_I(k_0)$, at four different times. It is obtained by numerically solving the time-dependent Schr\"odinger equation associated with Hamiltonian $\mathcal{H}$ using a disorder strength $W=2E_\zeta$ and an initial wave number $k_0\zeta=3.2$. The numerically-computed mesoscopic parameters are $\tau_s \approx 0.3\tau_\zeta$, $\ell_s \approx \zeta$, $k_0\ell_s \approx 3$, $\tau_B \approx 55\tau_\zeta$, $\ell_B\approx 165\zeta$ and $\xi \approx 55\zeta$. The diffuse momentum distribution consists of a broad background with a FWHM $\Delta k_D \sim 2/\ell_s$ ($\Delta k\zeta \approx 2$ here), reflecting the spectral broadening of the distribution due to disorder, and of a double peak structure emerging in the course of time.  The progressive symmetrization of the broad background occurs over a time scale roughly set by the Boltzmann transport mean free time $\tau_B$. The CBS peak grows at $-k_0$ and can be clearly observed after a few $\tau_s$, i.e. well before $\tau_B$. The CFS peak at $k_0$ is however only seen roughly after the Heisenberg time $\tau_H$ ($\tau_H \approx 40\tau_\zeta$ here), when the system enters the localization regime. Its height becomes comparable to the background only after a much longer time. In the stationary limit, both peaks become mirror images of each other, with a height twice the background and a very sharp width set by the localization length $\Delta k \sim 1/\xi$. The dashed lines mark the positions $k = \pm k_0$.
  }
\end{figure*}

\section{Model Hamiltonian and momentum distribution}

\subsection{1D Hamitonian and its statistical properties}

We consider the 1D wave dynamics of a particle with mass $m$ as described by the Hamiltonian $\mathcal{H}=-\frac{\hbar^2}{2m}\, \partial_x^2+V(x)$, where $V(x)$ is a spatially-correlated disordered potential with Gaussian statistics. At time $t=0$, the (free) initial state $|\Phi_0\rangle$ of the particle is supposed to be a plane wave state $|k_0\rangle$ with wave vector $k_0$ and energy $E_0=\hbar^2k_0^2/(2m)$.

Without any loss of generality, we assume here $V(x)$ to have a vanishing disorder-averaged mean value $\overline{V(x)} = 0$ since any finite mean value can always be swallowed up by a redefinition of the origin of energies. Throughout the paper $\overline{(\cdot\cdot\cdot)}$ denotes the average of the quantity $(\cdot\cdot\cdot)$ over the disorder configurations. Because we assume the disorder to have Gaussian statistics, Wick's theorem applies and all $n$-point potential correlators with $n$ odd vanish, whereas all potential correlators with $n$ even break down into products of 2-point correlators. For simplicity we further choose here the spatial 2-point correlator to be a Gaussian function
\begin{equation}
\label{eq:correlator}
\overline{V(x)V(x')} = C_2(x-x') = W^2 \exp(-\frac{(x-x')^2}{2\zeta^2}),
\end{equation}
where $W$ is the disorder fluctuations strength and $\zeta$ is the disorder correlation length. Note that, for a bulk system, the 2-point correlator $C_2$ only depends on the relative spatial separation since disorder average restores translation invariance. From a numerical point of view, such a disordered potential with Gaussian statistics and Gaussian 2-point correlator is generated by drawing uncorrelated random variables on a discrete grid and convoluting afterwards by a Gaussian function. The correlation length $\zeta$ defines a length scale, a time scale $\tau_\zeta=m\zeta^2/\hbar$ and an energy scale $E_\zeta=\hbar^2/(m\zeta^2)$ that we will use as the natural units of our system. One may note that our model does not cover the statistical properties of the speckle potential~\cite{kuhn2007,goodman2007}, which is commonly used in experiments and which does not obey Gaussian statistics. However we have numerically checked that the behaviors and conclusions reported here are not substantially modified for the speckle potential.

In the following, we will also consider Gaussian random $\delta$-correlated potentials for which $C_2(x)=U^2\delta(x)$, $\delta(x)$ being the Dirac delta distribution. From a numerical point of view, we investigated this case by solving the 1D Anderson model on a lattice~\cite{anderson1958,mkk1983} for energies close to the band edges. Starting from our correlated potential defined by Eq.~\eqref{eq:correlator}, the $\delta$-correlated limit is also obtained for particle energies $E\ll E_\zeta$, or equivalently for $k_E\zeta\ll 1$, where $k_E=\sqrt{2mE}/\hbar$ is the wave number of the particle at energy $E$. In this case, $U^2=\sqrt{2\pi}\, \zeta \,W^2$. For later purposes, we define the dimensionless disorder parameter
\begin{align}
\label{eq:alpha}
\alpha=\frac{U\sqrt{k_E}}{E} = (4\pi)^{1/4} \, \frac{W}{E^{3/4}E_\zeta^{1/4}},
\end{align}
which appears as a small parameter in weak-disorder perturbative expansions~\cite{weakscatt,lugan2009}.

Note that throughout the paper, we will use the convention $\langle x | k \rangle = \exp{(\mri kx)}$. The resolution of identity then reads
\begin{align}
\label{eq:reso}
\mathbbm{1} = \int dx \, |x\rangle\langle x| = \int \frac{dk}{2\pi} \, |k\rangle\langle k|,
\end{align}
with the orthonormality conditions $\langle x|x'\rangle = \delta (x-x')$ and $\langle k|k'\rangle = 2\pi \, \delta (k-k')$.

\subsection{Time evolution of the disorder-averaged momentum distribution}
To extract the disorder-averaged momentum distribution as a function of time, we numerically compute the time-evolved wave function $|\Psi(t)\rangle=|\Phi(t)\rangle \, \Theta(t)$ where $|\Phi(t)\rangle=\Exp{-\mri Ht/\hbar}|\Phi_0\rangle$ and $\Theta(t)$ is the Heaviside step function. The average density operator $\overline{\rho(t)} = \overline{|\Phi(t)\rangle\langle\Phi(t)|}$ can be split into two components, the ballistic one $\rho_b(t) =  \overline{|\Phi(t)\rangle}\,\overline{\langle\Phi(t)|}$ and the diffuse one 
$\rho_D(t) =\overline{|\delta\Phi(t)\rangle\langle\delta\Phi(t)|}$ where $|\delta\Phi(t)\rangle = |\Phi(t)\rangle-\overline{|\Phi(t)\rangle}$. The ballistic component $\rho_b(t) $ represents the time evolution of the initial plane wave mode $|k_0\rangle$ which is emptied by successive scatterings. For weak disorder, its time decay is exponential, $n_b(k,t) = |\,\overline{\langle k | \Phi(t)}\,|^2\approx \Exp{-t/\tau_s}\delta(k-k_0)$, with a time scale set by the scattering mean free time $\tau_s\equiv\tau_s(E_0)$~\cite{kuhn2007}. The other component $\rho_D(t)$ represents all the other initially-empty modes which are being populated  by the successive scatterings. Thus, after a few $\tau_s$, the diffuse component becomes the dominant contribution to $\overline{\rho}(t)$ and will be the focus of the rest of the paper. The momentum distribution at sufficiently large times $t$ is then simply approximated by $n(k,t) = \langle k |\overline{\rho(t)} | k\rangle = \overline{|\langle k|\Phi(t)\rangle|^2}  = n_b(k,t) + n_D(k,t) \approx n_D(k,t) =\langle k |\rho_D(t)| k\rangle$.

Figure \ref{fig:nk_dynamics_full} gives the numerical time-evolved diffuse momentum distribution $n_D(k,t)$ at four different times for $k_0\zeta = 3.2$ and a relatively strong disorder $W= 2E_\zeta$. One clearly sees two sharp peaks developing on top of a broad background. The background becomes symmetrical with respect to $k=0$ in the course of time while the two peaks become progressively mirror images of each other. The peak at $-k_0$ is the CBS peak and the one at $k_0$ is the CFS one. In the long-time limit, the two peaks have each a width of the order of $1/\xi$, where $\xi$ is the localization length at energy $E_0$~\cite{karpiuk2012}. The background is instead associated with diffusive transport and starts to develops after the time $\tau_s$ with a 
width of the order of $2/\ell_s$, where $\ell_s=v\tau_s$ is the scattering mean free path and $v$ is the group velocity~\cite{cherroret2012,karpiuk2012} ($v\approx \hbar k_0/m$ for weak disorder). The isotropization process leading to the background symmetrization occurs after a time scale known as the transport mean free time~\cite{plisson2012}. Neglecting interference corrections, a rough estimate of this time scale is given by the Boltzmann transport time $\tau_B$~\cite{kuhn2007}. When scattering is isotropic, which is the case in the low-energy limit $k_0\zeta \ll 1$ where the correlated potential appears as $\delta$-correlated, one has $\tau_B = \tau_s$. However, when $k_0\zeta$ increases, scattering becomes more and more anisotropic and $\tau_B$ increases much faster than $\tau_s$. The two time scales then become well separated. In 1D systems, and at weak disorder, one has $\xi =2\ell_B$~\cite{lugan2009}, where $\ell_B= v\tau_B$ is the transport mean free path. As a consequence, by varying $\zeta$ or $k_0$, one can easily reach a situation where $\xi \gg \ell_s$ and where both the CBS and CFS peaks become much sharper and thus become more easily distinguishable from the broader background. This feature of correlated disorder should help any experimental observation of the twin peaks.

Because multiple scattering paths consist of, at least, two scattering centers, the CBS peak can be in principle observed after two $\tau_s$, that is well below $\tau_B$ for anisotropic scattering. This is confirmed by the momentum distribution at time $t = 16\tau_\zeta \approx 53\tau_s \approx 0.29\tau_B$ in Fig.~\ref{fig:nk_dynamics_full}. As also seen, the CBS peak narrows in the course of time as multiple scattering fully develops and more and more scattering orders contribute to the effect. When the dynamics reaches AL, the peak width stabilizes at $1/\xi$. 

\begin{figure}
  \includegraphics[width=0.49\textwidth]{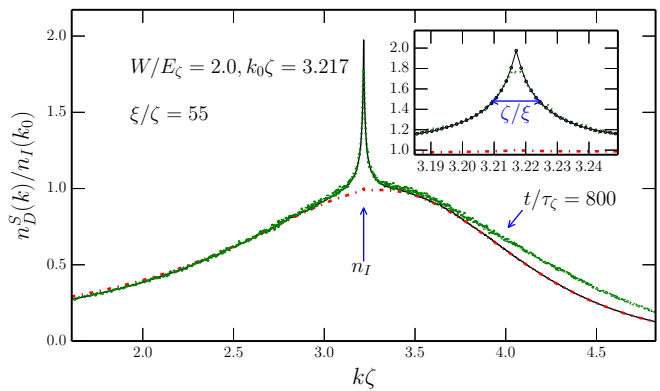}\caption{\label{fig:xi_RMT_E-1.6_W2z5} (Color online) The stationary momentum distribution $n_S(k)$ obtained in the long-time limit (black solid line) for the same parameters as in Fig.~\ref{fig:nk_dynamics_full}. The red dash-dotted line gives the incoherent background contribution $n_I(k)$ given by~\eqref{eq:diffusive_background}. For comparison, we also show the diffuse momentum distribution $n_D(k,t)$ at time $t =800\tau_\zeta$ (green dotted line). Inset: Zoom of the CFS peak observed at $k_0$ in $n_S(k)$. From the FWHM of the CFS peak, we get $\xi \approx 55\zeta$.
  }
\end{figure}

The dynamics of the CFS peak is a bit more subtle. It is absent in the early-time dynamics and starts to appear a bit below $\tau_B$, see the momentum distributions at times $t = 16\tau_\zeta \approx 0.29\tau_B$ and $t = 48\tau_\zeta \approx 0.87\tau_B$. Its time evolution involves the Heisenberg time $\tau_H=2\pi\hbar/\Delta$ where $\Delta$ is the mean level spacing associated with a localization box of size $\xi$~\cite{karpiuk2012}. Introducing $\nu\equiv \nu(E_0)$, the DOS at energy $E_0$, one has $\Delta =(\nu\xi)^{-1}$. In the weak-scattering limit $k_0\ell_s\gg1$, we have $\nu\approx 1/(\hbar\pi v)$ and $\tau_H$ boils down to the time needed to travel a localization length. At weak disorder, we thus expect $\tau_H \sim 2\tau_B$. It is important to note however that this estimate becomes bad when the disorder strength increases as $\tau_H$ becomes smaller than $\tau_B$. This is seen in Fig.\ref{fig:nk_dynamics_full} where we get the estimate $\tau_H \approx 0.7 \tau_B$. Remarkably, the measured CFS peak height first decreases in time, before increasing when the system enters the localized regime at time $t\sim\tau_H$, and finally saturates at a longer time scale. A hint at this behavior, shown in Fig.~\ref{fig:time_CFS_E-1.6W2z5}, can be found in the momentum distribution at time $t=320\tau_\zeta \approx 5.7\tau_B$ in Fig.\ref{fig:nk_dynamics_full} where the CBS peak already almost culminates at twice the background value while the CFS peak is still below this maximum contrast. The saturation of the CFS contrast occurs only after several $\tau_H$, as seen in Fig.~\ref{fig:xi_RMT_E-1.6_W2z5} obtained at time $t= 800\tau_\zeta \approx 14.3\tau_B$. In fact, as will be seen in the following Sections, the logarithmic repulsion between the energy levels induces an algebraic time dependence and thus a slow convergence dynamics of the CFS peak height to its maximum value.


\subsection{Relation to the eigensystem of the Hamiltonian}
Our target is to explain four main features of the momentum distribution in terms of the spectral properties of the random Hamiltonian: (1) the diffuse background, (2) the width of the CFS peak, (3) the height of the CFS peak and (4) the characteristic time scale of the CFS peak evolution. To this end we relate the momentum distribution at time $t$ to the eigensystem of the random bulk Hamiltonian $\mathcal{H}$. The latter consists of a dense pure-point spectrum $\{\varepsilon_\alpha\}$ and spatially exponentially-decaying eigenstates $\{|\varphi_\alpha\rangle\}$. From a numerical point of view, we consider a finite-size system of length $L$ with periodic boundary conditions, compute for each configuration the discrete eigenspectrum $\{\varepsilon_n, |\varphi_n\rangle\}$ of the discretized version $\mathcal{H}_L$ of the bulk Hamiltonian $\mathcal{H}$ and eventually take the limit $L\to\infty$. The discretization step in momentum space is $2\pi/L$ and $\Delta x$ in real space, such that $\mathcal{H}_L$ is represented by a $N\times N$ matrix with $N=L/\Delta x$. The eigenstates are normalized according to $\sum_{a=1}^{N} |\varphi_n(x_a)|^2 =1/\Delta x$ and $\sum_{a=1}^{N} |\varphi_n(k_a) |^2=L$ where $\varphi_n(x) = \langle x |\varphi_n\rangle$ and $\varphi_n(k) = \langle k |\varphi_n\rangle$ are Fourier transforms of each other. Starting from our initial plane wave state, we find
\begin{subequations}
\label{eq:momentum_spectrum_general}
\begin{align}
 &|\Phi(t)\rangle = \lim_{L\to\infty}\sum_n  \frac{\varphi^*_n(k_0)}{\sqrt{L}} \ \Exp{-\mri \varepsilon_n t/\hbar} \ |\varphi_n\rangle,\\
 &n_b(k,t) =\lim_{L\to\infty} \frac{1}{L} \left|\overline{\sum_n\varphi^*_n(k_0) \varphi_n(k) \Exp{-\mri\varepsilon_n t /\hbar}}\, \right|^2, \\
 &n(k,t) = \langle k| \overline{\rho(t)}| k\rangle \nonumber = \overline{\,|\langle k|\Phi(t)\rangle \,|^2}\\
 &= \lim_{L\to\infty} \frac{1}{L} \overline{\sum_{n,m}\varphi_n(k)\varphi^*_m(k)\varphi_m(k_0) \varphi^*_n(k_0) \Exp{-\mri(\varepsilon_n-\varepsilon_m)t/\hbar}}.\label{eq:momentum_spectrum_general_b}
\end{align}
\end{subequations}
As pointed out previously, one has $n(k,t) \approx n_D(k,t)$ as soon as $t \gg \tau_s$. Using Eq.~\eqref{eq:reso}, particle number conservation in momentum space reads
\begin{align}
{\rm Tr}\overline{\rho(t)} = \int \frac{dk}{2\pi} \, n(k,t) =1.
\end{align}

In the next Sections, we will analyze the properties of the momentum distribution through the looking-glass of the statistical properties of the eigensystem of the Hamiltonian $\mathcal{H}$ in momentum space.

\section{Momentum distribution in the long-time limit}

The momentum distribution in Eq.~\eqref{eq:momentum_spectrum_general_b} splits naturally into two components $n(k,t)=n_S(k)+\Delta n(k,t)$. The first term $n_S(k)$ is the stationary momentum distribution and is obtained for $\varepsilon_n =\varepsilon_m$. For our finite size simulations, there are no degeneracies in the eigenspectrum and $n_S(k)$ is simply obtained as the diagonal contribution $n=m$ of the summation. It thus reads
\begin{align}
\label{eq:stationary_distribution}
  n_S(k) = \lim_{L\to\infty} \frac{1}{L} \overline{\sum_n |\varphi_n(k)|^2|\varphi_n(k_0)|^2}.
\end{align}
The remaining time-dependent term $\Delta n(k,t)$ is then simply the off-diagonal contribution $n\not= m$ of the summation. Since almost degenerate eigenstates, with (small) energy difference $\varepsilon=\hbar\omega$, are spatially separated by a large distance of the order of $-\xi\ln|\omega\tau_H|$~\cite{mott_davis1979,sivan1987}, their momentum components become more and more uncorrelated as $\omega \to 0$. As a consequence, we indeed expect that $\lim_{t\to\infty}\Delta n(k,t) = 0$, confirming that the stationary momentum distribution is simply $n_S(k)$. Remembering that $n(k,t)$ is normalized to $1$ at all times, we find that $n_S(k)$ must also fulfill particle number conservation $\int dk \, n_S(k)/(2\pi)=1$. As a consequence $\Delta n(k,t)$ averages to zero at all times, $\int dk \, \Delta n(k,t)/(2\pi)=0$.

Since the disordered potential is real, our Hamiltonian is time-reversal symmetric and the spatial amplitudes $\varphi_n(x) = \langle x |\varphi_n\rangle$ can be chosen real. The Fourier components in momentum space thus satisfy $\varphi_n(k)=\varphi^*_n(-k)$. It is now straightforward to see from Eq.~\eqref{eq:stationary_distribution} that $n_S(k)$ is even in $k$ and thus symmetric with respect to $k=0$, as observed in the numerical simulations. This argument also shows that the CBS and CFS peak are perfect mirror images of each other in the stationary limit and thus must have the same width and height.

\subsection{Incoherent background contribution}
Eq.~\eqref{eq:stationary_distribution} involves the ensemble average of the sum of products $\overline{|\varphi_n(k)|^2\,|\varphi_n(k_0)|^2}$ where each Fourier amplitude  $\varphi_n(q)$ ($q=k,k_0$) appears as a random variable drawn from a statistical ensemble with well-defined statistical properties.
This remark invites the decomposition of $n_S(k)$ into an incoherent background contribution $n_I(k)$ and a coherent one $n_C(k) = n_S(k)-n_I(k)$. To this end, and omitting the limit $L\to\infty$ for brevity, we first write
\begin{align}
&n_S(k) =  \int \frac{dE}{L} \, \overline{\sum_n \delta(E-\varepsilon_n) |\varphi_n(k)|^2|\varphi_n(k_0)|^2}\nonumber \\
&= \int \frac{dE}{L} \, \overline{\sum_{n,m} \delta(E-\varepsilon_n) |\varphi_n(k)|^2|\varphi_m(k_0)|^2 \delta_{nm}}\nonumber \\
&= \int \frac{dE}{L^2\nu(E)} \overline{ \sum_{n,m} \delta(E-\varepsilon_n)|\varphi_n(k)|^2 \delta(E-\varepsilon_m)|\varphi_m(k_0)|^2},
\label{eq:ns(k)}
\end{align}
where we have used the prescription
\begin{align}
L\delta_{nm} \to \delta(\varepsilon_n-\varepsilon_m)/\nu(\varepsilon_n),
\end{align}
valid in the bulk limit and after disorder average. We now see that the stationary component has been recast under the form
\begin{equation}
n_S(k) =  \int \frac{dE}{2\pi} \, \frac{\overline{A(k,E) A(k_0,E)}}{2\pi \nu(E)},
\end{equation}
where
\begin{equation}
A(k,E) = \lim_{L\to\infty} \frac{2\pi}{L} \sum_n \delta(E-\varepsilon_n) |\varphi_n(k)|^2
\end{equation}
is the spectral function associated to the bulk retarded Green's function at energy $E$ for a given disorder configuration
\begin{align}
\overline{G^R_E} =& \overline{(E-\mathcal{H}+\mri 0^{+})^{-1}}.
\end{align}
The incoherent background contribution is then simply
\begin{align}
\label{eq:diffusive_background}
  n_I(k) =   \int \frac{dE}{2\pi} \, \frac{\mathcal{A}(k,E) \, \mathcal{A}(k_0,E)}{2\pi \nu(E)} 
\end{align}
with the disorder-averaged spectral function
\begin{align}
\mathcal{A}(q,E) &= \overline{A(q,E)} = -2 \, {\rm Im}\overline{G^R_E}(q) \nonumber \\
&= \lim_{L\to\infty} \frac{2\pi}{L}  \overline{\sum_n \delta(E-\varepsilon_n) |\varphi_n(q)|^2}.
\end{align}
The disorder-averaged DOS~\cite{kuhn2007} is obtained through
\begin{equation}
\label{eq:dos}
\nu(E) = \int \frac{dq}{(2\pi)^2} \, \mathcal{A}(q,E) = \lim_{L\to\infty} \frac{1}{L}  \overline{\sum_n \delta(E-\varepsilon_n)}.
\end{equation}
As one can see, the diffuse background is obtained by decoupling the components associated to different momenta, as if they were independent random variables. The same result can be derived from a diagrammatic perturbation theory, within the Boltzmann approximation~\cite{kuhn2007}. In the weak-disorder limit, the spectral function $\mathcal{A}(k,E)$ is a Lorentzian and $n_D(k)$ is just the convolution of two such Lorentzians. As a result, $n_S(k)$ has a relative FWHM of $2/\ell_s$~\cite{cherroret2012}. Notably, as seen in Figs.~\ref{fig:nk_dynamics_full} and \ref{fig:xi_RMT_E-1.6_W2z5}, and also noted in~\cite{lee2014a}, $n_D(k)$ is not peaked at $\pm k_0$ but at slightly shifted higher and lower $k$-values because of the weighting function $\nu(E)$ in the denominator of the integrand of Eq.~\eqref{eq:diffusive_background}.

Finally, using Eq.~\eqref{eq:dos}, one may note that~\cite{kuhn2007}
\begin{equation}
\int \frac{dk}{2\pi} \, n_I(k) = \int \frac{dE}{2\pi} \, \mathcal{A}(k_0,E) =1
\end{equation}
so that the incoherent background contribution satisfies particle number conservation. As a consequence, since $n_S(k)$ is also normalized to one, the coherent contribution must average to zero,
\begin{equation}
\int \frac{dk}{2\pi} \, n_C(k) =0.
\end{equation}


\subsection{Coherent contribution}
The coherent contribution to the stationary momentum distribution can also be recast as an integral over energies. Factoring out the incoherent background at energy $E$, we have
\begin{align}
  n_C(k) = \int \frac{dE}{2\pi} \frac{\mathcal{A}(k,E)\mathcal{A}(k_0,E)}{2\pi\nu(E)} \, C(E,k,k_0),
\end{align}
where $C(E,k,k_0)$ is the dimensionless auto-correlator of the fluctuations of the spectral function in momentum space at energy $E$,
\begin{align}
C(E,k,k_0) = \frac{\overline{\delta A(k,E) \delta A(k_0,E)}}{\mathcal{A}(k,E) \mathcal{A}(k_0,E)},
 \label{eq:momentum_wave_functions}
\end{align}
with $\delta A(k,E) = A(k,E) - \mathcal{A}(k,E)$. It is interesting to note the close connection between localization and the intensity correlations in momentum space of the wave functions: in the absence of these correlations, one simply gets $C(E,k,k_0)=0$ and $n_S(k) = n_I(k)$.

\begin{figure}
  \includegraphics[width=0.49\textwidth]{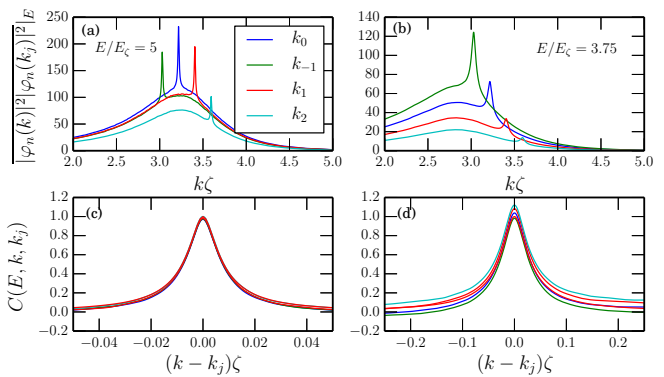}\caption{\label{fig:nk_shift_W2z5} (Color online) Top panels: momentum correlation function $\overline{|\varphi_n(k)|^2|\varphi_n(k_j)|^2}\big|_E$ (in arbitrary units) as a function of $k\zeta$ for different $k_j\zeta=3.2+0.2j$ ($j=0, \pm1,2$). Bottom panels: $C(E,k,k_j)$ as a function of $(k-k_j)\zeta$. The left panels correspond to $E=5E_\zeta$, $\xi \approx 61\ell_s \approx 55\zeta$. The right panels correspond to $E=3.75E_\zeta$, $\xi \approx 15\ell_s \approx 12\zeta$. Fo all panels, the disorder strength is $W=2E_\zeta$, the system size is $L=20000\zeta$ and the number of disorder realizations is $N_d=1000$. At $E=5E_\zeta$, $C(E,k,k_j)$ peaks at 1, is centered at $k_j\zeta$ and its shape is the same regardless of the actual value of $k_j$. At $E=3.75E_\zeta$, one can nevertheless see the effect of higher-order correlators as the wings of the curves now start to depends on the actual value of $k_j$.
}
\end{figure}

\begin{figure}
  \includegraphics[width=0.49\textwidth]{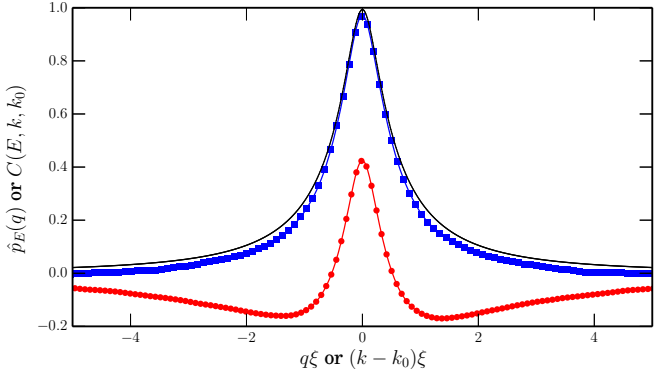}\caption{\label{fig:gogolin_momentum} (Color online) Comparison between the dimensionless intensity auto-correlator $C(E,k,k_0)$, computed numerically using Eq.~\eqref{eq:momentum_wave_functions} (blue squares connected by a continuous line to guide the eye), and the Fourier transform $\hat{p}_E(q)$ of the intensity auto-correlator in real space $p_E(x)$, computed theoretically using Eqs.~\eqref{eq:gogolin_formula} and \eqref{eq:gogolin_formula_fourier} (black solid line). The parameters are the same as those of the left panel of Fig.~\ref{fig:nk_shift_W2z5}, namely $W=2E_\zeta$, $k_0\zeta=3.2$, $\xi\approx=74\ell_s = 66\zeta$. Note that here $\xi \approx 1.2\, \xi_{{\rm RGF}}$. The figure also shows the data obtained for a $\delta$-correlated potential (red disks) with $\xi=2\ell_s$, $k_0\xi=178$ and $\alpha=0.14$, Eq.~\eqref{eq:alpha}. As one can see, whereas $C(E,k,k_0)$ peaks at 1 for the correlated potential ($\xi \gg \ell_s$), it only peaks at about 0.42 for the $\delta$-correlated one ($\xi \sim \ell_s$). 
   }
\end{figure}

\begin{figure}
  \includegraphics[width=0.49\textwidth]{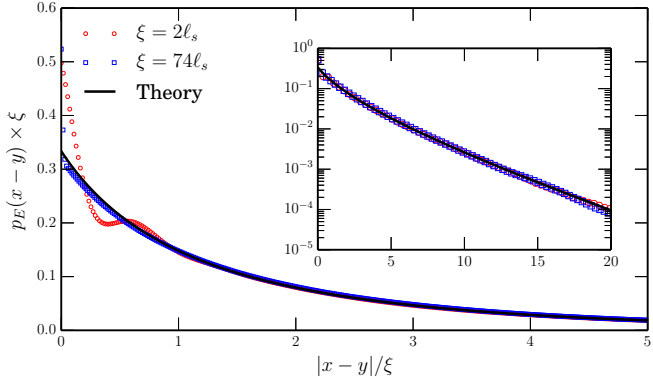}\caption{\label{fig:gogolin_spatial} (Color online) The intensity auto-correlator in real space $p_E(x)$ as a function of $x/\xi$. The black solid line gives the theoretical curve~\eqref{eq:gogolin_formula}, while the blue squares are numerical data extracted from the eigenstates of the random Hamiltonian with correlated disorder using Eq.~\eqref{eq:p_E_x}. The parameters used are the same as those for the left panel of Fig.~\ref{fig:nk_shift_W2z5}, namely $E= 5E_\zeta$, $W=2E_\zeta$ and $\xi= 74 \ell_s = 66\zeta \approx 1.2\xi$. The figure also shows the data obtained for the $\delta$-correlated potential (red circles) with $\alpha=0.14$, Eq.~\eqref{eq:alpha}. In this case, $\xi \approx 2\ell_s$. As one can see, the larger the ratio $\xi/\ell_s$, the closer we get to the theoretical prediction Eq.~\eqref{eq:gogolin_formula}. The oscillations observed at small $x$ in the $\delta$-correlated case come from the spatial correlation of the eigenstates over a distance of $\ell_s$. The inset shows the same curves but in log-scale for the $y$-axis.
}
\end{figure}


Let us now define the dimensionless auto-correlator
\begin{align}
p_E(x-y) = L \frac{\overline{\sum_n \delta(E-\varepsilon_n) |\varphi_n(x)|^2 |\varphi_n(y)|^2 }}{\nu(E)}
  \label{eq:p_E_x}
\end{align}
and its Fourier transform $ \hat{p}_E(q)=\int_{-\infty}^\infty\;dx\;\Exp{-\mri qx}p_E(x)$. Following the rationale of the previous section, it is easy to show that
\begin{align}
p_E(x-y) = \frac{\overline{A(x,E)A(y,E)}}{\nu^2(E)},
\end{align}
where $A(x,E) = \sum_n \delta(E-\varepsilon_n) |\varphi_n(x)|^2$ is the local DOS (LDOS), with disorder-average $\nu(E)$.
Then, in the regime where $\xi(E)\gg\ell_s(E)$, we numerically find that 
\begin{align}
C_E(k,k_0) &\approx \, \hat{p}_E(k-k_0)
\end{align}
in the CFS region $|k-k_0|\xi(E)\lesssim 1$, while
\begin{align}
C_E(k,k_0) &\approx \, \hat{p}_E(k+k_0)
\end{align}
in the CBS region $|k+k_0|\xi(E)\lesssim 1$, see Fig.~\ref{fig:nk_shift_W2z5} and \ref{fig:gogolin_momentum}. We have further checked that our numerically-extracted auto-correlator $p_E(x)$ agrees with the theoretical prediction~\cite{gogolin_1_1976,gogolin_2_1976,sanchezPalencia2007,sanchezPalencia2011}
\begin{align}
  p_E(x) = \frac{\pi^2}{16\xi(E)} \int_0^\infty & du \, \frac{u(1+u^2)^2\sinh \pi u}{(1+\cosh \pi u)^2} \times \nonumber\\
\times & \, \Exp{-\frac{(1+u^2)|x|}{4\xi(E)}}\label{eq:gogolin_formula},
\end{align}
see Fig.~\ref{fig:gogolin_spatial}. One has
\begin{align}
  p_E(x) &\approx \frac{1}{3\xi(E)}\Exp{-|x|/\xi(E)} &|x|\ll\xi(E), \\
  p_E(x)  &\approx \frac{\xi_E^{1/2}\pi^{7/2}}{32|x|^{3/2}}\Exp{-|x|/(4\xi_E)} &|x|\gg\xi(E).
\end{align}
The Fourier transform reads
\begin{align}
\label{eq:gogolin_formula_fourier}
\hat{p}_E(q) = \frac{\pi^2}{2} \int_0^\infty & du\,\frac{u(1+u^2)^3 \sinh \pi u}{(1+\cosh \pi u)^2} \times \nonumber\\
\times & \, \frac{1}{(1+u^2)^2+4q^2\xi^2(E)}.
\end{align}

By fitting our numerical results with either Eq.~\eqref{eq:gogolin_formula} or Eq.~\eqref{eq:gogolin_formula_fourier}, one can extract the localization length $\xi(E)$ for different values of the disorder parameters. The localization length can also be computed more efficiently by using another method, the recursive Green's function (RGF) one~\cite{lee2013_a}. We have checked that both methods give the same results when $\ell_s(E)>\zeta$.
However, the two estimates can be different when $\ell_s(E)\lesssim\zeta$: for example, $\xi(E)\approx1.2\,\xi_{\rm RGF}(E)$ for the data presented in Fig.~\ref{fig:nk_shift_W2z5}. Unless explicitly stated, the numerical values given for the localization length in this work will always refer to those obtained with the RGF method.

One can easily check that
\begin{align}
 &\hat{p}_E(q=0) = \int p_E(x) \, dx \nonumber \\
 & = \frac{\pi^2}{2} \int_0^\infty \frac{u (1+u^2)\sinh \pi u}{(1+\cosh \pi u)^2}  \, du= 1.
\end{align}
The immediate consequence of this result is that, when $\xi(E) \gg \ell_s(E)$, $n_C(\pm k_0) = n_I(k_0)$, so that $n_S(\pm k_0) = 2n_I(k_0)$. In the long-time limit, the CBS and CFS peak heights are thus exactly twice the incoherent background value as seen in Fig.~\ref{fig:xi_RMT_E-1.6_W2z5}.

\begin{figure}
  \includegraphics[width=0.49\textwidth]{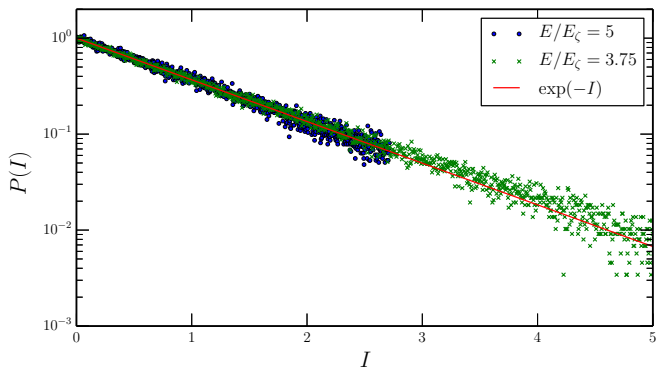}\caption{\label{fig:hist_psik} (Color online) The normalized intensity probability distribution in momentum space $P(I)=\overline{\delta(I-I(k,E))}$ computed at two different energies for a system size $L= 20000\zeta$ and a disorder strength $W=2E_\zeta$ (see text). With the chosen parameters, we have roughly $150-200$ eigenstates in a single disorder realization. The number of realizations is $N_d=1000$ and the size of the energy box used to compute the histogram is $\varepsilon = E_\zeta/16$. The solid red line gives the Poisson distribution.
}
\end{figure}

\begin{figure}
  \includegraphics[width=0.49\textwidth]{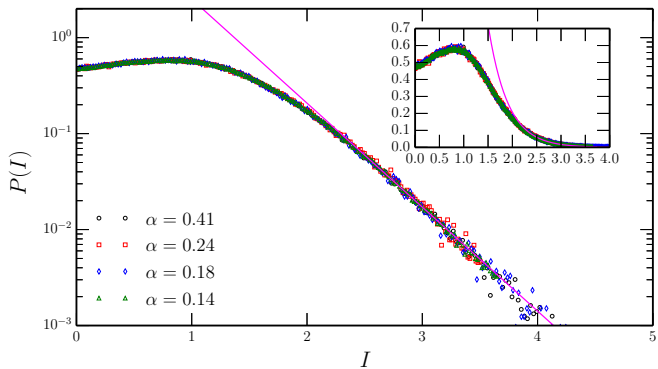}\caption{\label{fig:hist_psik_W0.25z0} (Color online) The normalized intensity probability distribution in momentum space $P(I)=\overline{\delta(I-I(k,E))}$ for a $\delta$-correlated random potential. Data sets obtained for different disorder parameters $\alpha$, Eq.~\eqref{eq:alpha}, collapse onto the same universal curve. The inset shows the same data but in linear scale. The solid magenta line is a fit of the distribution tail ($I>2.5$) by the exponential function $g(I)=a\,\Exp{-I/b}$ ($a = 31\pm1$ and $b=0.4\pm0.002$). The distribution $P(I)$ has a mean square value of $\overline{I^2}=1.42$, implying a CBS and CFS contrasts of $0.42$ instead of the value $1$ obtained for the correlated potential when $\xi \gg \ell_s$. Note that this value in fact depends on the momentum chosen to perform the computation.  }
\end{figure}

\begin{figure}
  \includegraphics[width=0.49\textwidth]{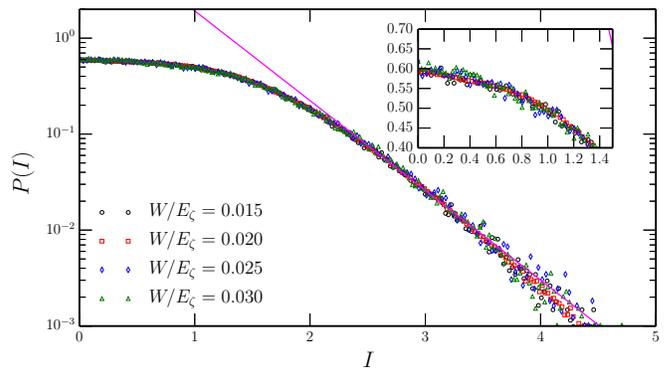}\caption{\label{fig:hist_psik_xi_over_ls2.45} (Color online) The normalized intensity probability distribution in momentum space $P(I)=\overline{\delta(I-\I(k,E))}$ (log-scale) for different disorder strengths $W$ of the correlated potential. The inset gives the same curve in linear scale. The energy is fixed at $E=0.1E_\zeta$. The on-shell momentum is then $k_E\zeta=0.45$ (see text). The ratio $\xi/\ell_s$ lies between $2.4$ to $2.5$, a bit larger than for the $\delta$-correlated case where $\xi = 2\ell_s$. The solid magenta line is a fit of the distribution tail ($I>2.5$) by the exponential function $g(I)=a\,\Exp{-I/b}$ ($a=16.6\pm0.6$, $b= 0.464\pm0.003$). The distribution $P(I)$ has a mean square value $\overline{I^2}=1.5$, implying a CBS and CFS contrast of $0.5$. Note that this value in fact depends on the momentum chosen to perform the computation. We anticipate that by increasing the ratio $\xi/\ell_s$, $\overline{I^2}$ increases towards its maximal value 2 irrespective of the chosen momentum.
  }
\end{figure}

\begin{figure}
  \includegraphics[width=0.49\textwidth]{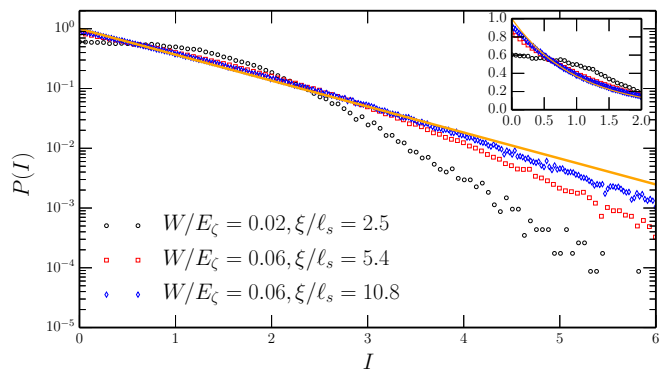}\caption{\label{fig:hist_psik_vary_xi_over_ls} (Color online) The normalized intensity probability distribution in momentum space $P(I)=\overline{\delta(I-I(k,E))}$ (log-scale) for different ratio $\xi/\ell_s$. The orange solid line gives the Poisson distribution. As one can see, the larger the ratio $\xi/\ell_s$, the closer $P(I)$ gets to the Poisson distribution. The inset gives the same curve in linear scale.
  }
\end{figure}

\subsection{Probability distribution in momentum space}

To better understand this peak height to background ratio of 2, we investigate the normalized probability distribution $P(I)$ of eigenfunctions in momentum space. This is done by computing the eigenstates within a small energy interval $[E-\varepsilon/2, E+\varepsilon/2]$ for a given disorder realization and then by constructing the histogram for the modulus square of these eigenstates at a chosen momentum $k$. The reduced intensity random variable is then $I(k,E) = A(k,E)/\mathcal{A}(k,E)$ and we find that $P(I)$ is given by the Poisson distribution
\begin{align}
  P(I)=\overline{\delta\bigl(I-I(k,E)\bigr)}=\Exp{-I}, \label{eq:momentum_poisson}
\end{align}
see Fig.~\ref{fig:hist_psik}. By contrast, similar studies~\cite{brody1981,berry1991,fyodorov1994,mirlin2000} have revealed that the probability distribution $Q(I)$ of real-space eigenfunctions for small system sizes $L\ll\xi$ is given by the Porter-Thomas distribution,
\begin{align}
  Q(I)=\overline{\delta\bigl(I-I(x,E)\bigr)}=\frac{1}{\sqrt{2\pi I}}\Exp{-I/2}\label{eq:real_potter}
\end{align}
where $I(x,E) = A(x,E)/\nu(E)$ is the reduced real-space intensity random variable. For large system sizes $L\gg\xi$, $Q(I)$ is dominated by rare events~\cite{altshuler1989,uski2000} and reads
\begin{align}
  Q(I)\sim\frac{1}{I}\Exp{-2I\xi/L}. \label{eq:real_localized}
\end{align}
We have duly checked that our real space numerical data indeed follow the predictions Eqs.~\eqref{eq:real_potter} and \eqref{eq:real_localized}.

The difference between Eqs.~\eqref{eq:momentum_poisson} and \eqref{eq:real_potter} can be explained by the number of random variables needed to describe the wave functions. Our Hamiltonian being time reversal invariant, $\varphi_n(x)$ is real with possible sign fluctuations, providing thus only one random variable to play with in real space. On the other hand, $\varphi_n(k)$ being complex, both its real and imaginary parts fluctuate independently, providing thus two random variables to play with in momentum space. If we assume that all these random variables obey a Gaussian statistics, then Eqs.~\eqref{eq:momentum_poisson} and \eqref{eq:real_potter} immediately follow. Since $C(E,k,k_0) = \overline{I(k,E)I(k_0,E)}-1$, it follows from Eq.~\eqref{eq:momentum_poisson} that
\begin{align}
  C_E(k_0,\pm k_0) = \int_0^\infty I^2P(I) \, dI-1=1.
\end{align}

\subsection{Ergodic picture}
To be more concrete, we now attempt to quantify the validity of the previous ergodic picture, where the phase of $\varphi_n(k)$ is assumed to be uniformly distributed over the interval $[0,2\pi]$. To this end, following the procedure explained above, we compute $P(I)$ for different values of $\xi/\ell_s$ obtained by varying $E$, $W$ and $\zeta$. To facilitate discussions, we restrict our investigation to the on-shell momentum $k=k_E = \sqrt{2m E}/\hbar$. The reason is that, when $k_E\ell_s\gg1$ (weak scattering regime), the spectral function $\mathcal{A}(k,E)$ is sharply-peaked around $k=k_E$.

We first consider $\delta$-correlated potentials. In this case, scattering is isotropic and $\tau_s$ and $\ell_s$ are the only relevant time and length scales of the problem, e.g. $\xi=2\ell_s$. Fig.~\ref{fig:hist_psik_W0.25z0} shows that all distributions $P(I)$ computed for different values of the dimensionless disorder parameter $\alpha$, see Eq.~\eqref{eq:alpha}, collapse onto the same universal curve with an average value $\overline{I^2}=1.42$ lower than 2. Note however that this value depends on the momentum chosen to do the computation: we would find another value if we had chosen $k\not= k_E$. This means that the CBS and CFS peaks are strictly smaller than the background. As one can immediately see, $P(I)$ clearly departs from the Poisson distribution for $\delta$-correlated potentials.

Going back to our correlated potential, we observe a similar behavior when $\xi$ is not too large compared to $\ell_s$, see Fig.~\ref{fig:hist_psik_xi_over_ls2.45} where all data still collapse onto a same universal curve and where $P(I)$ still departs from the Poisson distribution. However, as the ratio $\xi/\ell_s$ increases, the data keep collapsing onto a same universal curve but the distribution $P(I)$ now increasingly resembles the Poisson distribution, see Fig.~\ref{fig:hist_psik_vary_xi_over_ls} where the change in $P(I)$ is shown as $\xi/\ell_s$ increases. When the Poisson limit is reached, $\overline{I^2}=2$ irrespective of the actual value for the momentum. We numerically find that $P(I)$ is well described by the Poisson distribution at small intensities $I<5$ when $\xi$ and $\ell_s$ differ by an order of magnitude.

\begin{figure}
  \includegraphics[width=0.49\textwidth]{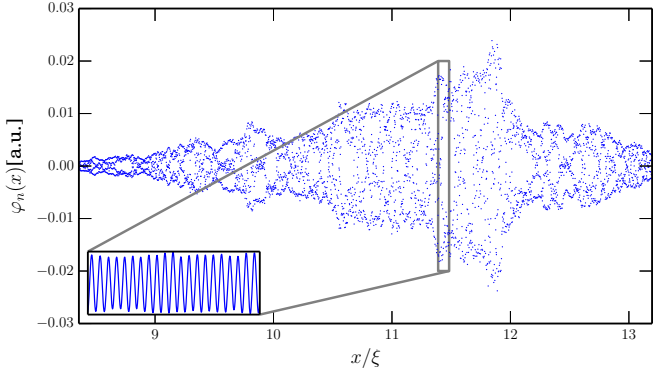}\caption{\label{fig:eigen_E-1.9W0.1z3.46ntot100000} (Color online) A typical eigenstate $\varphi_n(x)$ obtained in real space with a spatially-correlated disorder potential. The disorder parameters are $W=0.06E_\zeta$, $E=0.6E_\zeta$, $k_E\ell_s\approx134$, $k_E\xi\approx1440$, $L\approx22\xi$ and $\xi\approx11\ell_s$. The inset shows a zoom-in of the wave function over an interval of length $\ell_s$.
  }
\end{figure}

\begin{figure}
  \includegraphics[width=0.49\textwidth]{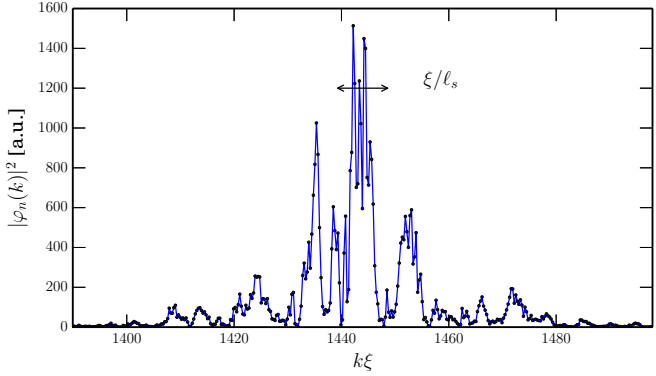}\caption{\label{fig:eigenk_E-1.9W0.1z3.46ntot100000} (Color online) The Fourier spectrum $|\varphi_n(k)|^2$ associated to the eigenfunction $\varphi_n(x)$ shown in Fig.~\ref{fig:eigen_E-1.9W0.1z3.46ntot100000}.
  }
\end{figure}

\begin{figure}
  \includegraphics[width=0.49\textwidth]{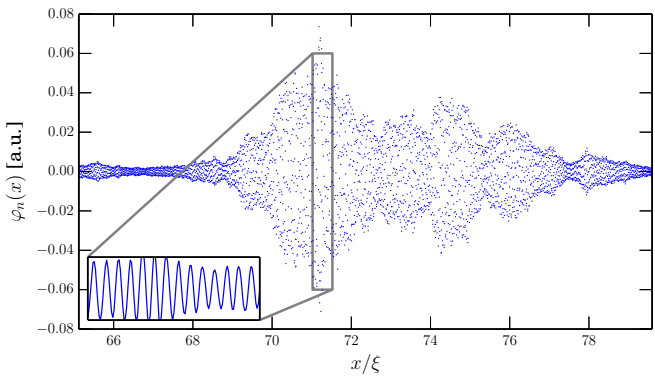}\caption{\label{fig:eigenx_E-1.6W0.25z0} (Color online) A typical spatial eigenstate $\varphi_n(x)$ obtained in real space for a $\delta$-correlated disorder potential. The disorder parameters are $\alpha=0.14$, $k_E\ell_s\approx89$, $k_E\xi\approx178$, $L\approx145\xi$ and $\xi = 2\ell_s$. The inset shows a zoom-in of the wave function over an interval of length $\ell_s$. This wave function looks similar to the one obtained in Fig.~\ref{fig:eigen_E-1.9W0.1z3.46ntot100000} with a correlated disorder potential
  }
\end{figure}

\begin{figure}
  \includegraphics[width=0.49\textwidth]{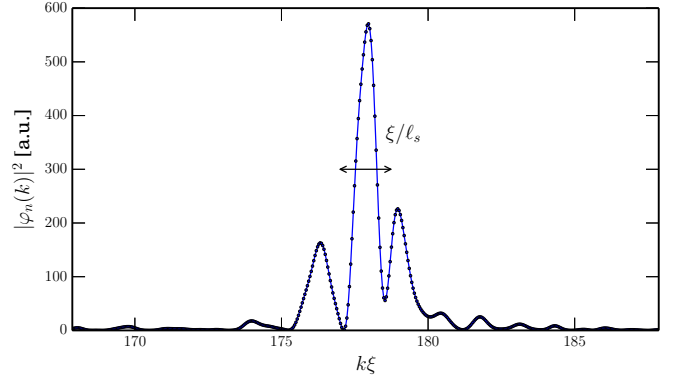}\caption{\label{fig:eigenk_E-1.6W0.25z0} (Color online) The Fourier spectrum $|\varphi_n(k)|^2$ associated to the eigenfunction $\varphi_n(x)$ shown in Fig.~\ref{fig:eigenx_E-1.6W0.25z0}. This Fourier spectrum is markedly different from the one obtained in Fig.~\ref{fig:eigenk_E-1.9W0.1z3.46ntot100000} with a correlated disorder potential as it looks smoother.
  }
\end{figure}

These observations can be understood by writing the eigenfunction $\varphi_n(x)$ in real space as~~\cite{fyodorov1994,mirlin2000}
\begin{align}
  \varphi_n(x) = a_n(x) \Phi_n(x) = a_n(x)\cos[k_Ex+\alpha_n(x)].
\end{align}
where $a_n(x)$ is a smooth envelop multiplying a rapidly-oscillating carrier $\Phi_n(x)$. In analogy with the problem of a wave propagating through a potential barrier, the phase $\alpha_n(x)$ suffers a random kick after each scattering. As a result, we expect $\alpha_n(x)$ to vary on a scale set by $\ell_s$ and the quickly-fluctuating component to be short-range correlated with the same scale $\ell_s$,
\begin{align}
  \overline{\Phi_n(x)\Phi_n(x')} \propto \Exp{-\frac{|x-x'|}{2\ell_s}},
\end{align}
see Figs.~\ref{fig:eigenx_E-1.6W0.25z0} and \ref{fig:eigen_E-1.9W0.1z3.46ntot100000}. On the other hand, $a_n(x)$ varies on a length scale set by the localization length $\xi$ and its statistics is determined by diffusion and localization effects. To obtain $\varphi_n(k)$, we have to Fourier transform $\varphi_n(x)$ over the system size $L$. Breaking the space integration over $L/\ell_s$ consecutive intervals of length $\ell_s$, we see that both the envelop $a_n(x)$ and the phase $\alpha_n(x)$ achieve almost constant (and random) values $a_{np}$ and $\alpha_{np}$ on each of these intervals labelled by $1\leq p\leq L/\ell_s$. 
We thus have
\begin{align}
  \varphi_n(k) \sim \sum_{p=1}^{L/\ell_s} a_{np}\Exp{\mri\alpha_{np}}.
  \label{eq:psik_sum}
\end{align}
It is however important to note that the number of terms that contribute significantly to the sum in Eq.~\eqref{eq:psik_sum} is proportional to $\xi/\ell_s$ as the spatial extension of $A(x)$ is set by the localization length. When $\xi/\ell_s\gg1$, we can appeal to the central-limit theorem and simply approximate $\varphi_n(k)$ by a complex number made of two independent Gaussian-distributed variables. These considerations justify the ergodic picture of the eigenfunctions in momentum space as long as $\xi/\ell_s \gg1$. In turn we can see why the ergodic picture does not apply to $\delta$-correlated potentials: as $\xi/\ell_s$ is small, the sum in Eq.~\eqref{eq:psik_sum} only contains a few terms and no simple limiting distribution can be inferred. In other words, when $\xi \gg \ell_s$, the wave functions suffer many random phase kicks and multiple scattering can efficiently scramble the phase of $\varphi_n(k)$, validating the ergodic picture, whereas it is not the case when $\xi\sim\ell_s$ where there are too few scattering events to efficiently scramble the phase. 
This can be clearly seen when comparing Fig.~\ref{fig:eigen_E-1.9W0.1z3.46ntot100000} and Fig.~\ref{fig:eigenk_E-1.9W0.1z3.46ntot100000} (correlated case) with Fig.~\ref{fig:eigenx_E-1.6W0.25z0} and Fig.~\ref{fig:eigenk_E-1.6W0.25z0} ($\delta$-correlated case). While the wave functions look (superficially) similar in real space, their Fourier spectra are markedly different, the first one looking more ``chaotic" than the second one.

\section{Time dependence of the CFS peak height}
\subsection{Relation to the DOS auto-correlator}
We now discuss the time dependent part $\Delta n_D(k_0,t)= n_D(k_0,t)-n_S(k_0) $ of the diffuse momentum distribution at the CFS momentum $k=k_0$. At sufficiently large times, $\Delta n_D(k_0,t) \approx  \Delta n(k_0,t) = n(k_0,t)-n_S(k_0)$ with
\begin{align}
\Delta n(k_0,t) = \frac{1}{L}\overline{\sum_{n\not= m}|\varphi_n(k_0)|^2 |\varphi_m(k_0)|^2 \, \Exp{-\mri(\varepsilon_n-\varepsilon_m)t/\hbar}}.
\end{align}
Writing
\begin{align}
\Delta n (k_0,t) = \int \frac{d\omega}{2\pi} \, \Exp{-\mri \omega t} \Delta\hat{n}(k_0,\omega),\label{eq:CFS_time1}
\end{align}
we have
\begin{align}
&\Delta \hat{n}(k_0,\omega) = \frac{2\pi\hbar}{L} \overline{\sum_{n\neq m} |\varphi_n(k_0)|^2 |\varphi_m(k_0)|^2\delta(\varepsilon+\varepsilon_n-\varepsilon_m)} \label{eq:phid_numerics}\\
&=\!\! \int \! \frac{2\pi\hbar dE }{L} \overline{\sum_{n\neq m} |\varphi_n(k_0)|^2 |\varphi_m(k_0)|^2\delta(E_{-}\!-\!\varepsilon_n)\delta(E_{+}\!-\!\varepsilon_m)},
\end{align}
where $E_{\pm} = E\pm \varepsilon/2$ and $\varepsilon= \hbar\omega$. We now note that, according to our numerical findings (not shown here), the fluctuations of the eigenfunctions decouple from the fluctuations of the level spacing between two eigenstates. Similar findings have been reported for the eigenfunctions in real space~\cite{altshuler1989}. Then:
\begin{align}
&\overline{\sum_{n\neq m} |\varphi_n(k_0)|^2 |\varphi_m(k_0)|^2\delta(E_{-}-\varepsilon_n)\delta(E_{+}-\varepsilon_m)} \approx \nonumber \\
&\frac{\mathcal{A}(k_0,E_+)\mathcal{A}(k_0,E_-)}{4\pi^2\nu(E_+)\nu(E_-)}\overline{ \sum_{n\neq m} \delta(E_{-}-\varepsilon_n)\delta(E_{+}-\varepsilon_m)},
\end{align}
leading to
\begin{align}
\Delta\hat{n}(k_0,\omega) = \hbar L \! \int \! \frac{dE}{2\pi}\,\mathcal{A}(k_0,E_{+})\mathcal{A}(k_0,E_{-}) \hat{K}_E(L,\omega),\label{eq:diffuse_propagator}
\end{align}
where
\begin{align}
  \hat{K}_E(L,\omega)= \frac{\overline{\delta \nu(E_{+}) \, \delta\nu(E_{-})}}{\nu(E_{+})\, \nu(E_{-})}
  \label{eq:dos_auto}
\end{align}
is the DOS auto-correlator with $\delta\nu(E) = \frac{1}{L}\sum_n\delta(E-\epsilon_n)-\nu(E)$ standing for the fluctuating part of the DOS. As one may note, it is an even function of $\omega$.



In the localized regime, considering that, the system of length $L$ can be broken into $L/\xi$ independent subsystems of length $\xi$, one arrives at the scaling relation~\cite{sivan1987,altland1995}
\begin{align}
\label{eq:KL}
  \hat{K}_E(L,\omega) \sim \frac{\xi(E)}{L} f(\omega\tau_H),
\end{align}
where $\tau_H = 2\pi\hbar /\Delta$ is the Heisenberg time associated with a localized subsystem, $\Delta=(\nu(E)\xi(E))^{-1}$ being the mean level spacing within this subsystem, and $f(\omega\tau_H)$ being the associated correlation function. The fact that $\lim_{L\to\infty}\hat{K}_E(L,\omega)=0$ shows that the eigenenergy spectrum in the localized regime follows a Poissonian statistics. However, in Eq.~\eqref{eq:CFS_time1}, we instead face the finite limit $\lim_{L\to\infty}L \hat{K}_E(L,\omega)=\xi(E) \, f(\omega\tau_H)$. This shows that the dynamics of the CFS peak is directly governed by the correlation function $f(\omega\tau_H)$.

It turns out that for $\omega\tau_H<1$ (equivalently $t>\tau_H$), the above picture of uncorrelated localized volumes is not sufficient. A more sophisticated model~\cite{altland1995} takes into account the exponential tail of the localized states which extends far beyond a single localized subsystem and states within different subsystems become coupled by tunneling~\cite{mott_davis1979}. These couplings give rise to correlations between the spectra of the different subsystems. For Hamiltonians belonging to the Gaussian Unitary Ensemble (GUE), it was predicted that $f(x)\propto \ln(x)$ in the range $\Exp{-L/\xi}<x\ll1$~\cite{altland1995}. Using supersymmetry methods, this logarithmic level repulsion was qualitatively understood as a consequence of the fact that localized states with an energy difference $\varepsilon = \hbar\omega$ are separated in space by a distance of $-\xi\ln|\omega\tau_H|$. Our numerical data here show that $f(x)$ follows this prediction even if our time-reversal symmetric Hamiltonian instead belongs to the Gaussian Orthogonal Ensemble (GOE). Up to our knowledge, we are not aware of a prediction for $f(x)$ in the GOE using supersymmetry tools.

\begin{figure}[htbp]
  \includegraphics[width=0.49\textwidth]{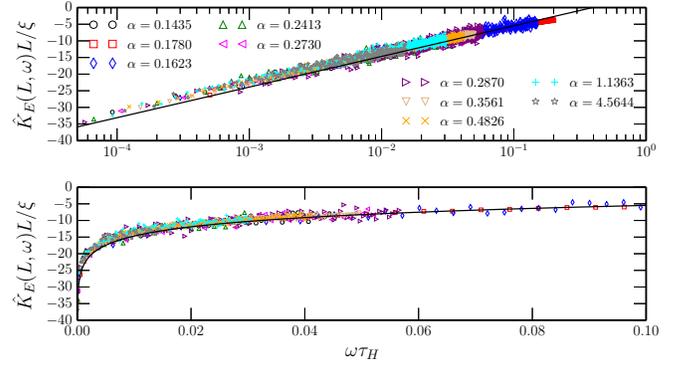}\caption{\label{fig:spacing_compareE_z0} (Color online) The DOS auto-correlation $\hat{K}_E(L,\omega)$ for $\delta$-correlated random potentials at different disorder strengths $\alpha$, Eq.~\eqref{eq:alpha}. The figure is a compilation of results obtained for system sizes ranging from $L=140 \xi$ to $L=600 \xi$. The number of disorder configurations used ranges from $N_d=10^5$ to $N_d=10^7$. The horizontal axis is shown in log (resp. linear) scale in the top (resp. bottom) panel. The black solid line gives the function $f(x)=2\beta\ln(\mu x)$ with $\beta=2$ and $\mu = 2.55$. 
  }
\end{figure}

\begin{figure}[htbp]
  \includegraphics[width=0.49\textwidth]{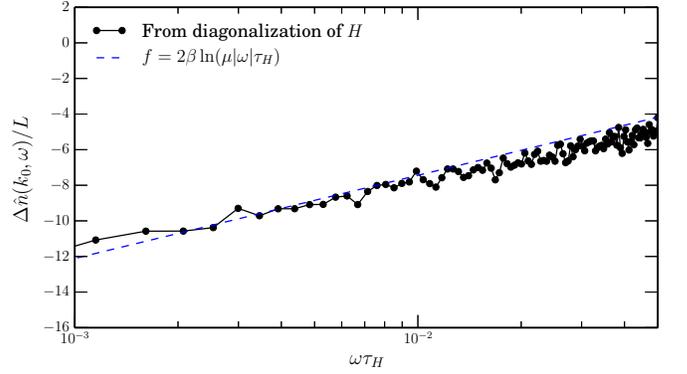}\caption{\label{fig:omega_fullE_E-1.6W0.25z0} (Color online) The intensity kernel $\Delta\hat{n}(k_0,\omega)$~\eqref{eq:diffuse_propagator} for a $\delta$-correlated potential with disorder strength $\alpha=0.1435$, Eq.~\eqref{eq:alpha}. The black continuous line gives the numerical data obtained by diagonalization of $H$ for a system size $L=73\xi(E_0)$, where $E_0= \hbar^2k^2_0/(2m)$ using the Anderson model on a lattice. The histogram has been constructed, for all pairs of different eigenstates, for $N_d=10^5$ disorder configurations. The blue dashed line gives $\Delta\hat{n}(k_0,\omega)$ computed with $f(x)=2\beta\ln(\mu x)$, where $\beta=2$ and $\mu = 2.55$.
 }
\end{figure}

%
%

\begin{figure}[htbp]
  \includegraphics[width=0.49\textwidth]{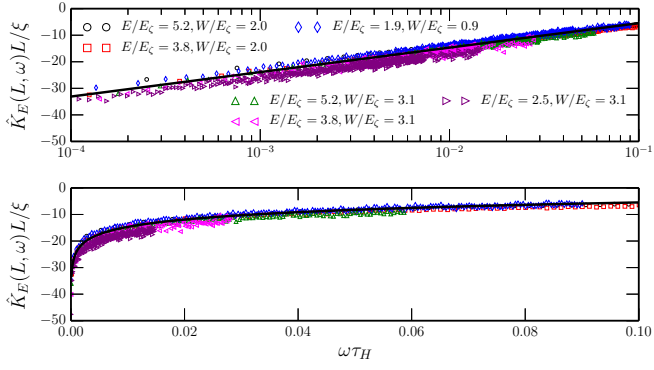}\caption{\label{fig:spacing_compareE_nonzero_zeta} (Color online) The DOS auto-correlation $\hat{K}_E(L,\omega)$ function for spatially-correlated random potential at different energies $E$ and disorder strengths $W$ (in units of the correlation energy $E_\zeta$). The figure is a compilation of results obtained for system sizes ranging from $L=70 \xi$ to $L=1400 \xi$. The number of disorder configurations used ranges from $N_d=10^6$ to $N_d=10^7$. The horizontal axis is displayed in log (resp. linear) scale in the top (resp. bottom) panel. The black solid line gives the function $f(x)=2\beta\ln(\mu x)$ with $\beta=2$ and $\mu =2.55$. One can see that the numerical data depart more from the analytical prediction at strong disorder $W>E$.
  }
\end{figure}

\subsection{Case of $\delta$-correlated potentials}

We now consider $\delta$-correlated random potentials. From the numerical diagonalization of $H$, and using $\xi$ calculated from the RGF method~\cite{lee2013_a} as a fitting parameter, we find that the following scaling function
\begin{align}
  f(\omega\tau_H) \equiv \frac{L \hat{K}_E(L,\omega)}{\xi(E)} = 2\beta \,  \ln(\mu |\omega|\tau_H),
  \label{eq:K_E_fitted}
\end{align}
where $\mu \approx 2.55$ and where $\beta = 2$ within a $5\%$ accuracy, fits well the data in the range $|\omega|\tau_H < 0.04$. This is consistent with the scaling $f(x) \propto \ln x$ provided $\omega$ is small enough (long-time limit). Fig.~\ref{fig:omega_fullE_E-1.6W0.25z0} shows the comparison between $f(x)$ computed using Eq.~\eqref{eq:phid_numerics} or using the theoretical prediction Eqs.~\eqref{eq:diffuse_propagator} with \eqref{eq:K_E_fitted}. as one can see, the agreement between the two methods is generally good, especially at small $\omega$. For sake of completeness, we also show in Fig.~\ref{fig:spacing_compareE_nonzero_zeta} that Eq.~\eqref{eq:K_E_fitted} is still a fair prediction for spatially-correlated potentials. We now use Eq.~\eqref{eq:K_E_fitted} to compute $\Delta n(k_0,t)$ and introduce the cutoffs $\pm 1/\tau_H$ for the integration over $\omega$ in Eq.~\eqref{eq:CFS_time1}. Writing $\Delta n(k_0,t) = \int \frac{dE}{2\pi} \, \Delta n_E(k_0,t)$, and assuming $E \gg \Delta$, we find
\begin{subequations}
\label{eq:CFS_time2}  
\begin{align}
  \Delta n(k_0,t) &= \int \frac{dE}{2\pi} \, \Delta n_E(k_0,t), \\
 \Delta n_E(k_0,t) &\approx -\frac{4\hbar}{\pi t} \mathcal{A}^2(k_0,E)\;\xi(E)\;{\rm Si}(t/\tau),\\
  & \approx -\frac{2\hbar}{t} \mathcal{A}^2(k_0,E)\;\xi(E),
 \end{align}
\end{subequations}
where $\tau = 2.55 \tau_H$. The last approximation is obtained in the long-time limit since the sine integral function ${\rm Si}(x)=\int_0^x dy \sin(y)/y \to\frac{\pi}{2}$ as $x\to\infty$. Writing now Eq.~\eqref{eq:diffusive_background} as $n_I(k_0)= \int \frac{dE}{2\pi} \, n_I(E,k_0)$, the CFS peak contrast at energy $E$, relative to its background value at same energy, then reads
\begin{align}
\label{eq:CFS_time3}
\mathcal{C}_E(k_0,t) = 1-\frac{\Delta n_E(k_0,t)}{n_I(E,k_0)} \approx 1 - \frac{2\tau_H(E)}{t}.
\end{align}
We thus find that the long-time dynamics $t\gg \tau_H$ of the CFS contrast is algebraic. From Eqs~\eqref{eq:CFS_time1} and \eqref{eq:diffuse_propagator}, and in the large time limit, we have
\begin{align}
\label{eq:CFS_time4}
\Delta n (k_0,t) = \int \frac{dE}{2\pi} \, \frac{\mathcal{A}^2(k_0,E)}{2\pi} \ \hbar L K_E(L,t),
\end{align}
where $K_E(L,t)$ is the Fourier transform of $\hat{K}_E(L,\omega)$. Comparison with Eq.~\eqref{eq:CFS_time2} show that
\begin{align}
\label{prediction}
\hbar \nu(E) L \,K_E(L,t) \approx - \frac{4 \tau_H}{\pi t} {\rm Si}(t/\tau).
\end{align}

\begin{figure}
  \includegraphics[width=0.45\textwidth]{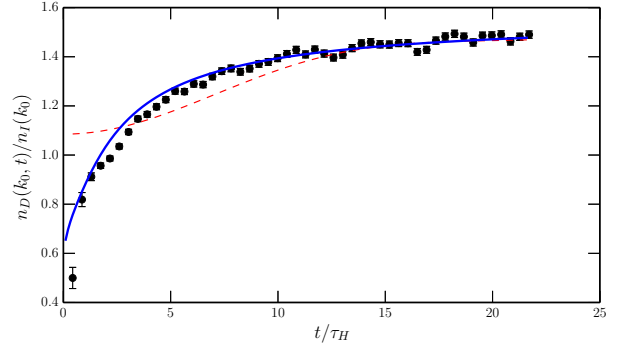}\caption{\label{fig:time_CFS_E-1.6W0.25z0} Time-dependent CFS peak height, $n_D(k_0,t)$ for $\delta$-correlated potentials with disorder strength $\alpha=0.1435$, Eq.~\eqref{eq:alpha}. The system size is $L=109\xi(E_0)$ and the scattering time is $\tau_s(E_0)=\tau_H(E_0)/4$, where $E_0=\hbar^2k^2_0/(2m)$. The black circles with error bars give the numerical data obtained by solving Schr\"odinger equation and an average over $N_d= 10^4$ disorder configurations. The red dashed line gives the prediction Eq.~\eqref{eq:CFS_time4} computed with Eq.~\eqref{prediction}. The blue solid line gives the theoretical prediction obtained by plugging the conjecture Eq.~\eqref{eq:MMA_CFS} into~Eq.\eqref{eq:CFS_time5}.
  }
\end{figure}

\begin{figure}
  \includegraphics[width=0.45\textwidth]{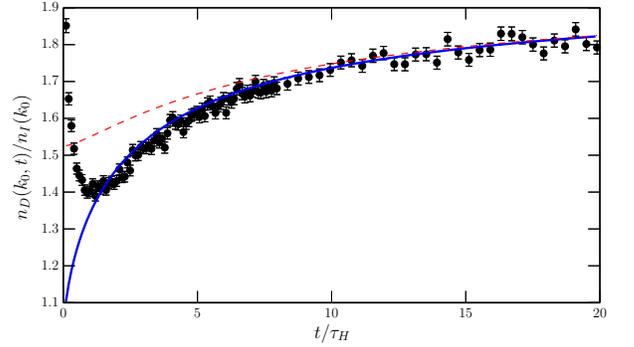}\caption{\label{fig:time_CFS_E-1.6W2z5}Time-dependent CFS peak, $n_D(k_0,t)$ for Gaussian correlated potentials with disorder strength $W=2E_\zeta$. The initial momentum is $k_0\zeta=3.2$. The black circles with error bars give the numerical data obtained by solving Schr\"odinger equation and an average over $N_d= 10^4$ disorder configurations. The system size is $L=90\xi(E_0)$, where $E_0=\hbar^2k^2_0/(2m) = 0.5(k_0\zeta)^2 E_\zeta = 5.12 E_\zeta$. The red dashed line gives the prediction Eq.~\eqref{eq:CFS_time4} computed with Eq.~\eqref{prediction}. The blue solid line gives the theoretical prediction obtained by plugging the conjecture Eq.~\eqref{eq:MMA_CFS} into~Eq.\eqref{eq:CFS_time5}.
}
\end{figure}

\begin{figure}
  \includegraphics[width=0.45\textwidth]{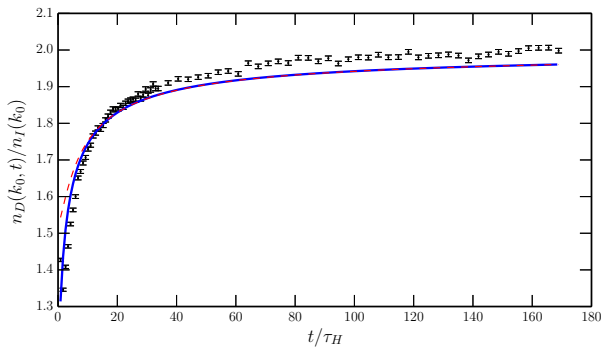}\caption{\label{fig:time_CFS_E-1.6W3z5}Time-dependent CFS peak $n_D(k_0,t)$ as a function of time $t$ (in units of the Heisenberg time $\tau_H$) for Gaussian correlated potentials. The disorder strength is $U=3E_\zeta$, the initial momentum is $k_0\zeta=3.2$ (corresponding to energy $E_0 =5.12 E_\zeta$) and the system size is $L=800 \xi(E_0)$. The black dots with error bars give the numerical data obtained by solving the Schr\"odinger equation and an average over $N_d=10^5$ disorder configurations. The red dashed line gives the prediction Eq.~\eqref{eq:CFS_time4} computed with Eq.~\eqref{prediction}. The blue solid line gives the theoretical prediction obtained by using the GUE prediction Eq.~\eqref{eq:MMA_CFS} to compute Eq.~\eqref{eq:CFS_time5}. The slight discrepancy observed at large times comes from a finite-size effect.
  }
\end{figure}

\begin{figure}
  \includegraphics[width=0.48\textwidth]{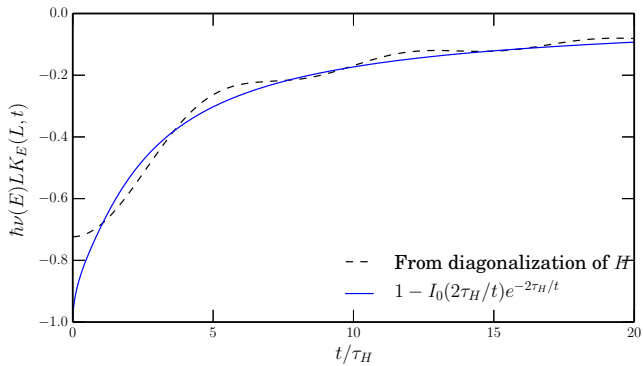}\caption{\label{fig:time_K_E_E-1.6W0.25z0} The black dashed line gives the function $\hbar \nu(E) L \,K_E(L,t)$ as a function of time $t$ (in units of the Heisenberg time $\tau_H$) for $\delta$-correlated potentials at disorder strength $\alpha=0.14$, Eq.~\eqref{eq:alpha}. The function $K_E(L,t)$ is the Fourier transform of the DOS auto-correlator $\hat{K}_E(L,\omega)$, Eq.~\eqref{eq:dos_auto}. The integration range is restricted to $|\omega|\tau_H\leq1$ and done using eigenstates within a small energy range $[E, E+\varepsilon]$, such that $\xi(E)$ and $\nu(E)$ vary by less than 2\% over this range. The number of disorder configurations used is $N_d=10^7$. The observed oscillation is due to the introduction of the cut-offs $\pm 1/\tau_H$. As one can see, the agreement with the GUE prediction of Ref.~\cite{micklitz2014}, Eq.~\eqref{eq:MMA_CFS} and \eqref{eq:CFS_time5}, is very good (blue continuous curve).
   }
\end{figure}

Fig.~\ref{fig:time_CFS_E-1.6W0.25z0} shows the comparison between \eqref{eq:CFS_time4} computed with Eq.~\eqref{prediction} and data obtained by numerically solving Schr\"odinger's equation with a $\delta$-correlated potential. Figs.~\ref{fig:time_CFS_E-1.6W2z5}-\ref{fig:time_CFS_E-1.6W3z5} show the same comparison for systems with spatially-correlated potentials. All data show that the CFS peak rises rapidly when the system enters the localized regime but then saturates only algebraically. As one can see, the agreement is good at large enough times.

It turns out that Eq.~\eqref{eq:CFS_time3} is exactly the asymptotics predicted at long times for the CFS peak observed in a quasi-1D system under a weak magnetic field (GUE symmetry class)~\cite{micklitz2014},
\begin{align}
 \mathcal{C}_E(t) &= I_0(2\tau_H/t)\Exp{-2\tau_H/t} \label{eq:MMA_CFS} \\
&\approx 1-\frac{2\tau_H}{t}+\mathcal{O}\left((2\tau_H/t)^2\right)\nonumber
\end{align}
In Figs.~\ref{fig:time_CFS_E-1.6W0.25z0}-\ref{fig:time_K_E_E-1.6W0.25z0}, we plot
\begin{align}
\label{eq:CFS_time5}
\Delta n(k_0,t) = \int \frac{dE}{2\pi} \, \frac{\mathcal{A}^2(k_0,E)}{2\pi\nu(E)} \, \big(1- \mathcal{C}_E(t) \big)
\end{align}
based on the full GUE prediction. Though relying on results derived for a GUE system while we deal with a GOE system, we see a good agreement with our data over a large range of times. A reason may be that the leading contribution to the CFS peak for the GUE system are diagrams made of ``maximally-crossed" ladders which are immune to time-reversal symmetry breaking~\cite{micklitz2014}. These very same diagrams are also at play in our GOE system~\cite{cherroret}. One can however see that the data for the correlated case in Fig.~\ref{fig:time_CFS_E-1.6W3z5} departs from the theoretical prediction at large enough times. We believe this is a finite-size effect. Indeed, strictly speaking, the CFS peak signals that the wave dynamics is bounded in space. For a bulk disordered system, the mechanism is AL. However for a finite-size system, even diffusion is bounded and turns out to contribute a CFS effect in momentum space, an effect related to the dynamical echo in real space~\cite{prigodin1994}. How to distinguish a CFS peak originating from bounded diffusion in a disordered box or from AL in disordered bulk systems, in particular for experimental purposes, will be addressed elsewhere.

\section{Conclusion}
In this paper, going beyond the diagrammatic analysis presented in~\cite{karpiuk2012}, we have performed a thorough statistical analysis of the eigensystem of a 1D Hamiltonian with a random potential. We have related the width, height and time-dependence of the CFS peak, appearing in the course of time in the momentum distribution of a quasi-monochromatic wave packet, to the correlations existing between the eigenstates of the system as well as between the DOS fluctuations. In particular, we have shown that the long-time dependence of the CFS peak originates from the logarithmic level repulsion between localized states with a time scale that is about twice of the Heisenberg time. Our results confirm that the spatial scale characterizing the CFS peak is proportional to the localization length, whereas the time scale governing its dynamics is the Heisenberg time. We believe that the experimental observation and study of the CFS peak in a 1D geometry is within the reach of current ultracold atom experiments. 

The recent theoretical work by Micklitz {\it et al.} revealed the robustness of the CFS peak in quasi-1D systems in the presence of a weak magnetic field~\cite{micklitz2014}. From this point of view, it would be interesting to consider other symmetry classes, and even higher dimensional systems, to quantify further the relationship between the CFS effect and Anderson localization in bulk systems. In particular, it would be interesting to address the 3D case where a metal-insulator transition is known to take place, with a mobility edge delineating extended states from localized ones.

\begin{acknowledgments}
The Authors wish to thank C. M\"uller, N. Cherroret and D. Delande for their interest in the work. KLL thanks A. Altland for helpful communications. The Centre for Quantum Technologies is a Research Centre of Excellence funded by the Ministry of Education and the National Research Foundation of Singapore.
\end{acknowledgments}

$ $\\

\bibliography{momentum_CFS_1D}

\begin{thebibliography}{55}
\bibitem [{wea()}]{weakscatt}%
  \BibitemOpen
  \href@noop {} {}\bibinfo {note} {Within the Born approximation, $\alpha
  \propto (k_E\ell_s)^{-1/2}$. Hence the weak-scattering regime $k_E\ell_s \ll
  1$ is equivalently characterized by the condition $\alpha \ll1$.}\BibitemShut
  {Stop}%
\end{thebibliography}

\end{document}